# On the use of near-neutral Backward Lyapunov Vectors to get reliable ensemble forecasts in coupled ocean-atmosphere systems

Stéphane Vannitsem*[1], and Wansuo Duan[2]

[1]*Royal Meteorological Institute of Belgium, Brussels, Belgium*

[2]*LASG, Institute of Atmospheric Physics, Chinese Academy of Sciences, Beijing 100029, China*

**ABSTRACT**

The use of coupled Backward Lyapunov Vectors (BLV) for ensemble forecast is demonstrated in a coupled ocean-atmosphere system of reduced order, the Modular Arbitrary Order Ocean-Atmosphere Model (MAOOAM). It is found that overall the best set of BLVs to initialize a (multiscale) coupled ocean-atmosphere forecasting system are the ones associated with near-neutral or slightly negative Lyapunov exponents. This unexpected result is related to the fact that these sets display larger projections on the ocean variables than the others, leading to an appropriate spread for the ocean, and at the same time a rapid transfer of these errors toward the most unstable BLVs affecting predominantly the atmosphere is experienced. The latter dynamics is a natural property of any generic perturbation in nonlinear chaotic dynamical systems, allowing for a reliable spread with the atmosphere too. Furthermore, this specific choice becomes even more crucial when the goal is the forecasting of low-frequency variability at annual and decadal time scales. The implications of these results for

---

*Corresponding author : Stéphane Vannitsem

Email: Stephane.Vannitsem@meteo.be

operational ensemble forecasts in coupled ocean-atmosphere systems are briefly discussed.

## 1. Introduction

An ensemble forecast is an operational procedure developed in the late twentieth century in order to take into account the amplification of uncertainties in the initial conditions and generate a set of potential future outcomes of the atmospheric dynamics (Toth and Kalnay, 1993; Molteni et al, 1996). This approach, originally based on theoretical considerations on probabilistic forecasts (Epstein, 1969), is now an essential component of any operational forecasting system aiming at providing information on the quality of the forecasts and/or warnings on possible unexpected and sometimes extreme events. In more recent years, additional sources of uncertainties were incorporated describing the presence of model errors, e.g. (Buizza et al, 1999; Buizza, 2019).

Operational ensemble forecasts were originally developed in the context of weather forecasts with a time horizon from one to two weeks. Rapidly, this method of uncertainty quantification also percolated in other fields of environmental and climate sciences, like for instance in Hydrology (e.g. Roulin and Vannitsem, 2005, and reference therein) or in climate projections (e.g. Tebaldi and Knutti, 2007).

A key desirable property of an ensemble forecasts is to be *reliable* or *calibrated*. An ensemble is said reliable or calibrated if the observation (or the reference) can be considered as a possible member of the ensemble statistically indistinguishable from any other forecast issued by the model, or in other words the probability distribution of the forecasts is statistically consistent with the observations (or the reference values). This is a joint property of the forecasts and the observations (Gneiting, et al 2007). The other key property is the property of *sharpness* which refers to the concentration of the probability distribution of the forecasts and is a property associated with the forecasts only (Gneiting et al, 2007), but will not be investigated here. Different methods have been proposed to check for the reliability of ensemble forecasts, and some important tools can be found in Wilks (2011). A first element that should be checked when evaluating the reliability of ensemble forecasts is to compare the mean square error between the ensemble mean and the observation, and the variance of the ensemble. If both quantities are close to each other, the variability of the ensemble members as described by its second moment appropriately represent the forecast uncertainty. It is then usually said that the ensemble is well calibrated (Leutbecher and Palmer, 2008).

Most of the ensemble forecasts produced in Meteorological Centers are not perfectly calibrated, but tuning the amplitude or pattern of the initial condition errors, or the model-uncertainty perturbations allows for getting better results at the space and time scales of interest (see e.g. Kalnay, 2003; Buizza et al, 2008). A similar tuning problem arises when dealing with ensemble forecasts of other climate components as discussed in Zanna et al (2019). For initial condition errors, perturbations were historically combinations of Singular Vectors or Bred Vectors. Nowadays it can also be combined with ensembles generated by data assimilation (Buizza et al, 2008).

When dealing with multiscale systems the problem becomes more difficult because the error dynamics can also evolve on different time scales as illustrated for instance in Vannitsem (2017). How to build ensemble forecasts providing reliable probability distributions of all the variables of the system is therefore a new challenge (Sandery and O'Kane 2014; O'Kane et al, 2019). Important efforts were devoted to the development of ensemble forecasts based on Bred modes tuned to describe the slow error growth on seasonal to decadal time scales for the ocean dynamics or the coupled ocean-atmosphere dynamics (Cai et al, 2003; Vikhliaev et al, 2007; Yang et al 2008, 2009; Frederiksen et al 2010; Baehr and Piontek, 2014; O'Kane et al, 2019). This tuning based on a rescaling at monthly timescales induces a saturation of the errors acting at short timescales as illustrated in an idealized context by Peña and Kalnay (2004) and Norwood et al (2013), and preserves the instability acting on longer time scales. These are therefore good candidates for simulating the uncertainty for long term forecasts at seasonal and decadal time scales. Building on these findings based on Bred vectors, one may wonder why ensemble forecasts targeting the dynamics at seasonal to decadal timescales is more appropriate with such unstable modes with low amplifications. The understanding of this feature is one of the main goals of the present work.

As the Bred modes are empirical modes that are affected by nonlinearities and highly dependent on the breeding time and amplitude, we investigate that problem using the Backward Lyapunov Vectors (BLVs) that are known to correspond to orthogonal Bred modes for small rescaling amplitudes (Feng et al, 2016; Duan and Huo, 2016). The BLVs are vectors that only depend on the background instantaneous fields, and not on any rescaling time and amplitudes. In this sense it makes them more appropriate tools to investigate the theoretical question on the link between timescales of the instabilities and reliability of ensemble forecasts. At the same time, this investigation allows for clarifying whether such modes can be appropriately used for initializing reliable coupled ocean-atmosphere ensembles.

After a brief description of the coupled (multiscale) ocean-atmosphere model used in the present paper (Section 2), the BLVs will be described, together with their main dynamical properties (Section 3). The experimental setup for the investigation of the impact of the choice of BLVs on the reliability of ensemble forecasts is performed in Section 4. Section 5 contains the main results, indicating that the most unstable BLVs are not the most appropriate fields to initialize a reliable ensemble forecasts, but rather the BLVs displaying a rather slow growth rate or decay. The reasons for this feature are further discussed in the concluding remarks of Section 6.

**2. The coupled ocean-atmosphere model**

Recently a reduced-order coupled ocean-atmosphere system has been developed allowing for extensive dynamical analyses. The equations of motion describing the dynamics are the quasi-geostrophic equations for a two-layer atmosphere and a one-layer ocean superimposed on an infinitely deep quiescent ocean layer (Vallis, 2006; Vannitsem, 2017). The temperature within the ocean is considered as a passive scalar transported by the ocean flow. The coupling between the ocean and the atmosphere is made through radiative, heat, and momentum transfers.

The solutions of these equations are expanded in Fourier series truncated severely at low wavenumber, and are plugged into the model equations. The resulting equations are then projected on the Fourier modes that are retained, leading to a set ordinary differential equations (De Cruz et al, 2016). The domain of definition of these fields is a rectangular domain with $0 \leq x \leq \frac{2\pi L}{n}$ and $0 \leq y \leq \pi L$ where n is the aspect ratio between the meridional and the zonal extents of the domain, and L the characteristic space scale. The boundary conditions for the atmosphere are periodic along the zonal direction and free-slip along the meridional direction (no flux through the boundaries along the meridional direction). For the ocean, a closed basin is imposed with no flux through the boundaries. The most advanced version of this model is freely available on Github at https://github.com/Climdyn/MAOOAM, in which additional information on its installation, the computer languages and the typical solutions that are generated are provided.

This model was found to display multiscale chaotic dynamics, with for some parameter values and resolutions, a low-frequency variability within the atmosphere reminiscent of the variability found in the real atmosphere at mid-latitudes (Vannitsem et al 2015; Vannitsem 2015; De Cruz et al 2016). This low-frequency variability is crucially dependent on the strength of the wind stress at the interface between the ocean and the atmosphere and the presence of an energy balance scheme between the two components. In the present work, the original version of the model developed by Vannitsem et al. (2015) will be used. In this version the four fields, the barotropic and baroclinic atmospheric streamfunctions, and the ocean streamfunction and temperature fields are given by

$$\psi = \sum_{i=1}^{10} \psi_{a,i} F_i,$$

$$\theta = \sum_{i=1}^{10} \theta_{a,i} F_i,$$

$$\Psi = \sum_{i=1}^{8} \Psi_{o,i} \varphi_i,$$

$$T = \sum_{i=1}^{8} T_{o,i} \varphi_i,$$

where $\psi$ and $\theta$ are the barotropic and baroclinic streamfunctions for the atmosphere; the $F_i$ are 10 Fourier modes (low wavenumbers) compatible with the boundary conditions of the equations for the atmospheric dynamics that are periodic in the zonal direction; the $\varphi_i$ are 8 Fourier modes compatible with the closed boundaries imposed to the ocean (no fluxes in both the zonal and meridional directions) and $\Psi$ and T are the streamfunction and temperature fields of the ocean dynamics.

Most of the parameter values used in the present work are the same as in Vannitsem (2017), except that the radiative input from the sun is now fixed to $C_0 = 350 \ W/m^2$ and the friction coefficients between the ocean and the atmosphere used in the two configurations discussed below are $C = 0.01 \ kg/(m^2 \ s)$ and $C = 0.016 \ kg/(m^2 \ s)$. For the other parameters, see Vannitsem (2017).

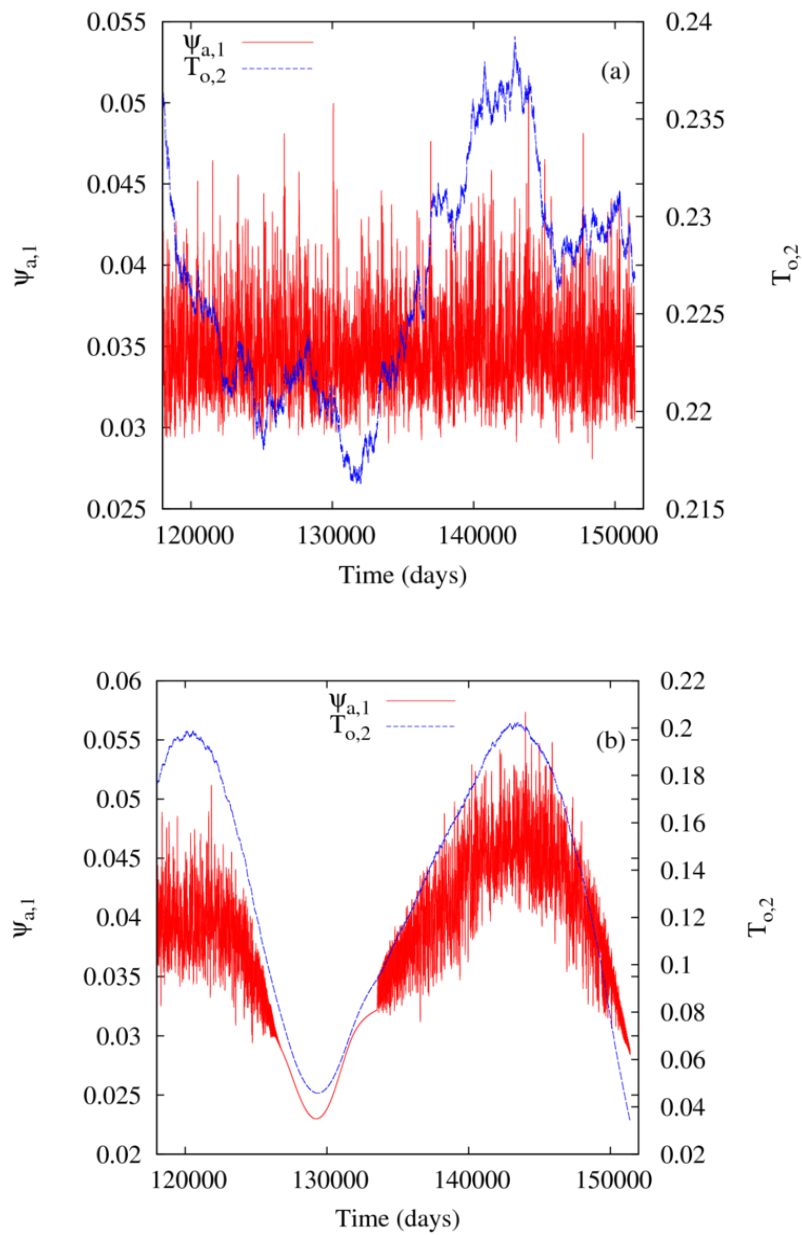

Figure 1: Time evolution of 2 key variables of the model, θ$_{a,1}$ and T$_{o,2}$, for (a) $C = 0.01\ kg/(m^2\ s)$ and (b) $C = 0.016\ kg/(m^2\ s)$.

Figure 1 displays the two typical time evolutions for (a) $C = 0.01\ kg/(m^2\ s)$ and (b) $C = 0.016\ kg/(m^2\ s)$. Two key variables are displayed, θ$_{a,1}$, the first mode of the barotropic atmospheric streamfunction, and T$_{o,2}$, the second mode of the ocean temperature field. In the first panel, a very erratic behavior is found for both the atmospheric and oceanic variables, while in the second a clear low-frequency signal is present for both variables on a time scale of the order 25,000 days (about 70 years). Obviously in the second configurations there are periods of very weak variability, while others of very intense variability. This variability is due to the stronger coupling with the ocean, inducing a bifurcation toward new qualitative solutions (see Vannitsem et al, 2015; Vannitsem, 2017). The low-frequency variability is not very realistic for the current evolution of the large-scale atmosphere at mid-latitudes, but is a way to mimic the presence of oscillations like the North-Atlantic Oscillation or the Southern Oscillation. It therefore allows in an idealized setting figure out what happens when qualitatively different periods are present in the dynamics.

**3. The Backward Lyapunov Vectors**

In the ergodic theory of dynamical systems, three types of (un)stable vectors (or fields) that are local properties of the flow are well defined and are known as Forward, Backward and Covariant Lyapunov vectors, FLVs, BLVs and CLVs, respectively. Let us introduce these vectors briefly.

Consider first a dynamical system described by the ordinary differential equation,

$$\frac{d\boldsymbol{x}}{dt} = \boldsymbol{f}(\boldsymbol{x}, \{\gamma\}, t) \quad (1)$$

Where $\boldsymbol{x}$ is the set of variables, $\{\gamma\}$ a set of parameters, and $t$ the time. In the following we will consider that there is no explicit dependence on time since our results are investigated in the context of an autonomous version of the coupled ocean-atmosphere system. These equations can be linearized to describe the evolution of infinitesimally small perturbations, $\delta \boldsymbol{x}$, as

$$\frac{d\delta \boldsymbol{x}}{dt} = \left.\frac{\partial \boldsymbol{f}}{\partial \boldsymbol{x}}\right|_{\boldsymbol{x}} \delta \boldsymbol{x} \quad (2)$$

and the solution of Eq. (2) can be formally written as

$$\delta \boldsymbol{x}(t) = \boldsymbol{M}(t, t_0) \delta \boldsymbol{x}(t_0) \quad (3)$$

where $\boldsymbol{M}(t, t_0)$ is the resolvent matrix describing the amplification of small perturbations. The Oseledets theorem tells us that in the limit of infinite positive time, the product $(\boldsymbol{M}(t, t_0)^T \boldsymbol{M}(t, t_0))^{1/(2(t-t_0))}$ has a well-defined limit and the logarithm of the eigenvalues of this asymptotic matrix are called the Lyapunov exponents, $\sigma_i$ (Oseledets, 2008). The eigenvectors of this matrix are called the Forward Lyapunov vectors, already denoted previously as FLVs (Vautard anf Legras, 1996). These FLVs are still dependent on $t_0$ and are therefore local properties at time $t_0$. Note that these vectors only depend on this single initial time. Similarly one can define another matrix, $(\boldsymbol{M}(t, t_0) \boldsymbol{M}^T(t, t_0))^{1/(2(t-t_0))}$, and when one takes the limit for $t_0$ going to infinite negative time, the asymptotic matrix is also well defined and similarly to the previous one, the logarithm of its eigenvalues are the Lyapunov exponents. Its eigenvectors are now called the Backward Lyapunov Vectors and are defined at time $t$. Note that the Lyapunov exponents are usually ranked in decreasing order, and the whole set of exponents is called the Lyapunov spectrum.

These vectors and their properties were extensively discussed in recent years in the literature, in particular with respect to the significance of the eigenvectors of the matrices above (Vautard and Legras 1996; Trevisan and Pancotti, 1998; Pazo et al 2008; Kuptsov and Parlitz 2012). Note that these vectors are not perturbations that are covariant under the dynamics of the error in the tangent (linearized) space of the system. The CLVs, denoted here as $\boldsymbol{g}_i(t)$, are characterized by an amplification in the tangent space of the trajectory of the form

$$M(t,t_0)\boldsymbol{g}_i(t_0) = \vartheta_i(t,t_0)\boldsymbol{g}_i(t) \tag{4}$$

where $\vartheta_i(t,t_0)$ is the stretching factor along the ith CLV and $M(t,t_0)$ the fundamental matrix -- see also Gaspard (1998) for a detailed discussion on the properties of the stretching rates and the fundamental matrix. These vectors are not necessarily orthogonal and are usually computed as the intersections of a succession of subspaces defined by the FLVs and BLVs, see Vautard and Legras (1996). Once an infinitesimally small perturbation is introduced along one of these vectors or a combination of them, it will stay in the subspace defined by the corresponding set of vectors along the trajectory.

Starting from the stretching rate one can also define the Lyapunov exponent as,

$$\sigma_i = \lim_{t \to \infty} \frac{1}{t} ln|\vartheta_i(t,t_0)| \tag{5}$$

On the other hand, if any perturbation has a whatever small component along the first CLV, or equivalently the first BLV, then it will rapidly grow in the course of positive time along the dominant instability of the system. This property is very important since if a perturbation is taken at random, then it will anyway converge to the dominant instability after a time typically associated with the difference of the successive exponents of the Lyapunov spectrum.

The BLVs also constitute the limits for infinitesimally small perturbations of the Bred vectors, properly orthogonalized. This correspondence makes the BLVs interesting candidates as perturbations for ensemble forecasts, with the advantage that it is not necessary to include parameters such as rescaling time and amplitude.

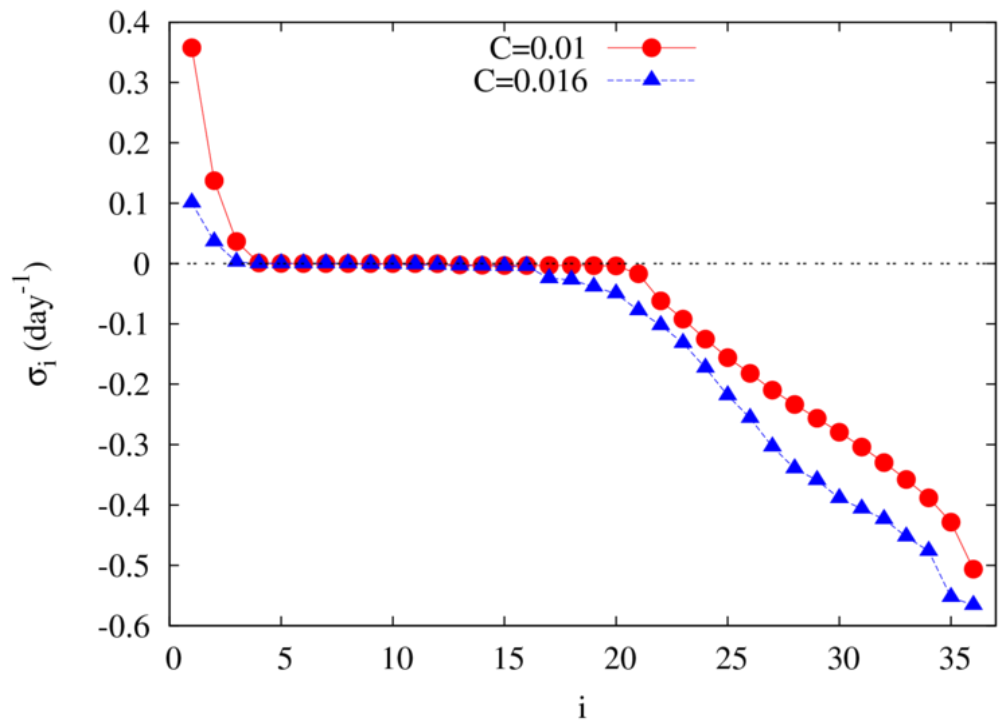

Figure 2: Lyapunov spectra for the solutions displayed in Figure 1 and generated using two different values of $C = 0.01 \ kg/(m^2 \ s)$ (full circles) and $C = 0.016 \ kg/(m^2 \ s)$ (full triangles).

The Lyapunov spectrum can be computed using standard algorithms as discussed for instance in Kalnay (2003) or Kuptsov and Parlitz (2012). They are displayed for the two parameter sets used in Figure 1, for $C = 0.01 \ kg/(m^2 \ s)$ and $C = 0.016 \ kg/$ ($m^2\ s$). A first important remark is that there are a few positive exponents, a set of exponents close to 0 (only one is exactly 0), and a set of negative exponents. As discussed extensively by Vannitsem and Lucarini (2016), the set of exponents close to 0, are associated with CLVs with a large projection on the ocean modes, while the very positive and very negative ones have very little projection on ocean modes. This suggests that the CLVs and associated exponents close to 0 are quantities describing the (slow) dynamics of errors related to the influence of the ocean. This has also been demonstrated by splitting explicitly the Lyapunov exponents and vectors of the ocean and the atmosphere in the same model by Penny et al (2019). The impact of vectors used to perturb the initial conditions in ensemble forecasts should therefore be highly dependent on the index of these vectors. The preferential projection of the vectors along the variables of the coupled model are also present for the BLVs. We will use these vectors in the following sections.

**4. Ensemble forecasts: Experimental setup**

To clarify the impact of specific choices of Lyapunov subspaces defined by the BLVs on the quality of ensemble forecasts, some idealized experiments will be performed in the context of the reduced-order coupled ocean atmosphere system introduced in Section 2. Experiments will be done for the solutions of the model displayed in Figure 1, assuming that there is no model error affecting the forecasts. In this case, twin forecasting experiments are performed in the following way:

- A long reference run is performed as displayed in Figure 1.
- For a set of N=1000 different initial conditions taken at random along this long run, ensemble forecasts are performed with M=20 ensemble members, including the control forecast.

- The initial condition error between the control forecast and the reference trajectory is sampled from a uniform distribution between [-5 $10^{-7}$, 5 $10^{-7}$] along all variables.

- The amplitude of the random perturbations around the initial conditions of control forecast is also sampled with the same uniform distribution and then projected along the subset of BLVs of interest. If the number of BLVs used is smaller than the total number of BLVs, S=36, then the amplitude of the perturbation will be smaller than the one of the original perturbation, as some components of the random perturbation are neglected. More importantly the orientation of the perturbation will not be isotropic anymore. This implies that the ensemble will be unreliable by construction as the perturbations introduced around the control forecast will be smaller than the one separating the reference and the control and will affect only certain specific directions in phase space.

- A final step can be made by tuning the amplitude of the initial perturbations along the subset of BLVs chosen in order to improve the reliability of the ensemble. A similar tuning step is used when initializing ensembles with Bred Vectors or Singular Vectors.

- The reliability is comparing the mean square error of the ensemble mean (MSE):

$$\text{MSE} = \frac{1}{N} \sum_{i=1}^{K} \|x_i - \overline{y}_i\|^2$$

where $\overline{y}_i$ is the mean of the ith ensemble forecast, $x_i$ the corresponding reference solution, N the number of different ensemble forecasts along the trajectory of the solution and $\|.\|$ the usual L2-norm; and the variance of the ensemble (SPREAD):

$$\text{SPREAD} = \frac{1}{N} \sum_{i=1}^{N} \frac{1}{K-1} \sum_{k=1}^{K} \|y_{k,i} - \bar{y}_i\|^2$$

where $K$ is the number of ensemble members. If both are equal, then the ensemble is considered in our setting as reliable.

- A second evaluation based on a proper scoring rule developed by Dawid and Sebastiani (1999) which provides an estimate of the quality of the first and second moments of the forecast distribution. This scoring rule is related to the ignorance score advocated as among the most useful scores for probabilistic forecasts (Roulston and Smith; 2002; Benedetti, 2010; Smith et al, 2015). Let us consider one element $j$ of the vectors $x_i$ and $y_{k,i}$ (with $i=1,…, N$ the index of the realization and $k=1,…, K$, the index of the ensemble member), the Dawid-Sebastiani Score (DSS) can be written as

$$DSS(x_{i,j}) = \frac{1}{2} \log(2\pi) + \frac{1}{2} \log \sigma_{i,j}^2 + \frac{1}{2} \frac{(K-3)}{(K-1)} (\bar{y}_{i,j} - x_{i,j})^2 / \sigma_{i,j}^2$$

where $\sigma_{i,j}^2$ is the variance estimator of the ensemble for variable $y_{k,i,j}$ with $k=1, …, K$. Corrections for the finite size of the ensemble can also be taken into account (Siegert et al, 2019; Leutbecher, 2019), but as we are comparing ensembles with equivalent number of members, this does not need to be taken into account. Furthermore, in the following analyses we will also drop the first term of DSS for the same reason. An average can then be performed over all $i =1, …, N$. The DSS can be extended to multivariate datasets (Dawid and Sebastiani, 1999), but the focus will be put here on specific single components.

As a reference test, the 36 BLVs – or equivalently, the original random perturbations – are used. In this case the ensemble should be perfectly reliable as the initial uncertainty between the control and the reference runs is sampled from the same distribution as the perturbations introduced in the ensemble forecasts. Figure 3 displays the MSE and SPREAD for the four fields of the model, the barotropic and baroclinic atmospheric streamfunctions and the streamfunction and temperature within the ocean. So 8 curves are displayed. The MSEs are represented with symbols and the SPREAD with lines. All curves of MSE and SPREAD are superimposed on each other, indicating that the ensemble forecasting system is reliable by design.

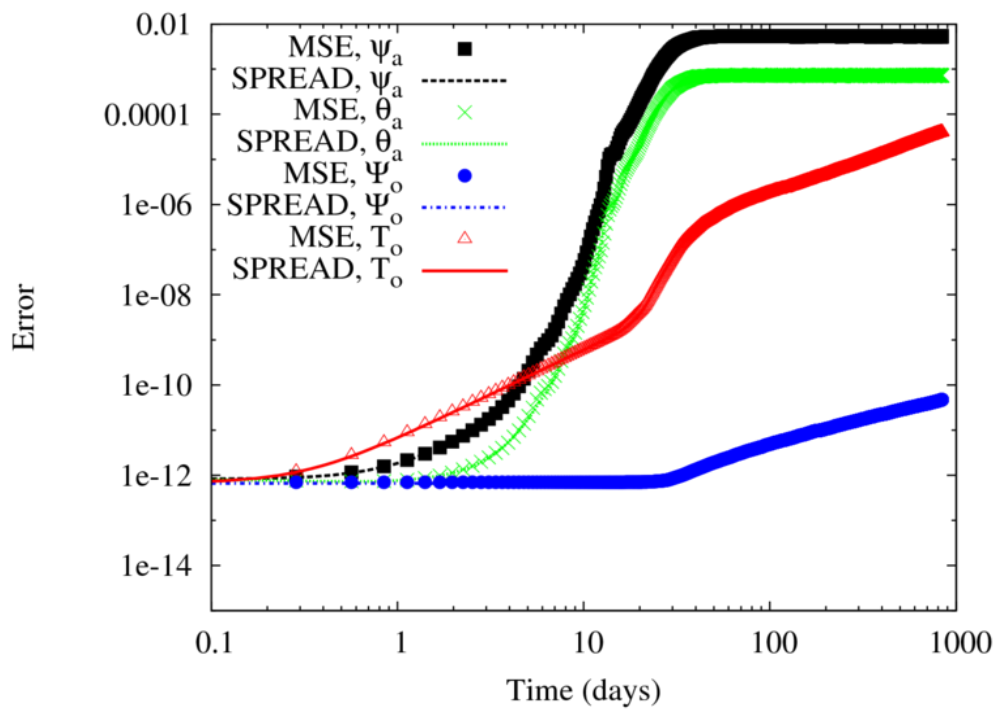

Figure 3: Mean Square Error of the ensemble mean (MSE, symbols) and variance of the ensemble (SPREAD, curves) as a function of lead time for the four fields of the coupled system, namely the barotropic atmospheric streamfunction (full squares, black dashed curve), the baroclinic atmospheric streamfunction (crosses, green continuous curve), the ocean streamfunction (full circles, blue dash-dotted curve), and the ocean temperature (open triangles, brown continuous curve). Perturbations introduced along the 36 BLVs of the system, with a perfectly reliable ensemble as MSE and SPREAD are superimposed on each other. The reference solution as in Fig. 1a.

In the following, we present the results on the reliability of ensemble forecasts based on several subsets of BLVs: (i) the 1st BLV associated with the dominant Lyapunov exponent; (ii) the 10 first BLVs, associated with the 10 first Lyapunov exponents corresponding to the most unstable directions (and part of the ones close to 0; (iii) the set of BLVs associated with the 11 to 20 Lyapunov exponents spanning a large portion of the spectrum close to 0; (iv) the set of BLVs associated with the 21 to 30 Lyapunov exponents corresponding to a set of weakly stable directions.

## 5. Results

Figure 4 shows the same quantities as in Fig. 3, MSE and SPREAD, when perturbations are made along specific BLVs. In panel (a) only the projection of the perturbation along the first BLV is used. The ensemble is clearly under-dispersive for all fields as the SPREAD is smaller than MSE. This should be expected as the variance of the random perturbation along the first BLV is smaller than the total perturbation. What is however very instructive is the fact that the SPREAD is smaller by several orders of magnitude for the streamfunction and temperature fields in the ocean. This underdispersion persists until more than 10 days for temperature and more than 50 days for the streamfunction within the ocean.

Tuning the amplitude of the perturbations along the different fields (or variables) of the first BLV can be performed in a way to improve the SPREAD of the ensemble. For instance, by increasing the amplitude of the perturbation with factors 2, 4, 20, 20 for the barotropic and baroclinic atmospheric streamfunctions and for the temperature and streamfunction within the ocean, respectively, the SPREAD of the barotropic and baroclinic atmospheric streamfunctions can be partially improved, while the ocean fields are still highly under-dispersive. If one increases further the amplitude of the perturbation to factors 2, 10, 200, 200, one can improve the baroclinic atmospheric streamfunction but the barotropic one is now degraded, with no improvement of the SPREAD of the ocean fields (Fig. 4f). So tuning is not allowing real improvement here.

In panel (b), the projections of the random perturbations along the BLVs from 1 to 10 are kept, with a better match between MSE and SPREAD for the atmospheric fields. But it is striking to note that the SPREAD for ocean temperature and streamfunction are still largely under-dispersive, even if all the unstable (and some of the stable) BLVs are used as perturbations. In panel (c) the use of the projections of the random perturbations along the BLVs from 11 to 20 shows however much better results with a SPREAD much closer to the MSE than in panels (a) and (b), although these vectors are associated with negative Lyapunov exponents. This is also true when perturbing along vectors 21 to 30 which are even more stable BLVs than the subset from 11 to 20. To complete the analysis, panel (e) shows the case with BLVs from 1 to 20 covering the most unstable and the near-zero negative Lyapunov exponents. This last experiment provides some improvements for the atmospheric fields as compared to the case with vectors associated with the near-zero exponents only, but no visible improvements for the oceanic fields. This point will be now taken up by investigating the DSS.

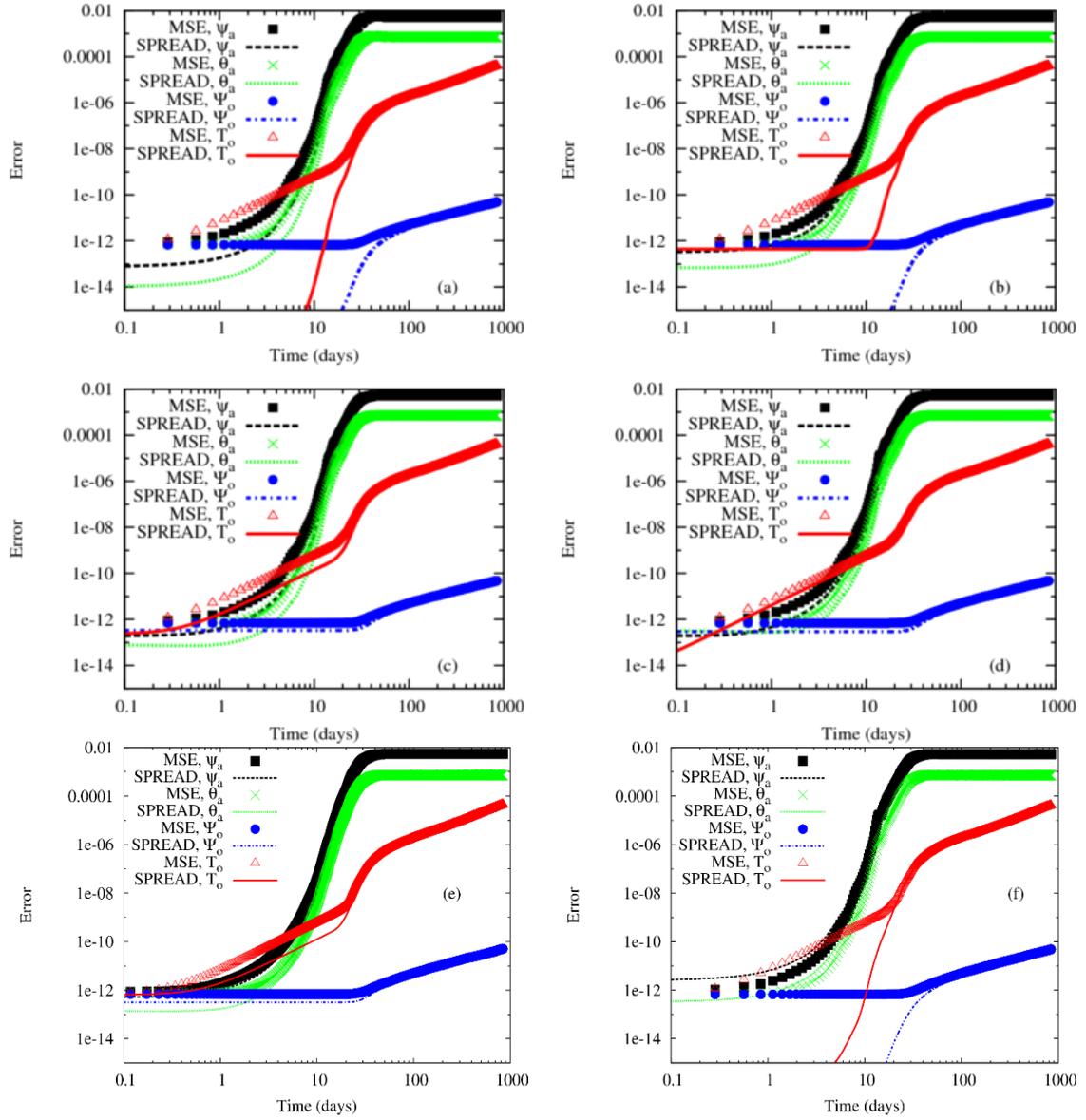

Figure 4: As in Fig. 3, but now the perturbations are limited to a set of BLVs: (a) the first BLV; (b) the 10 first BLVs; (c) BLVs from 11 to 20; (d) BLVs from 21 to 30, (e) BLVs from 1 to 20; and (f) the first BLVs with a rescaling of the set of variables.

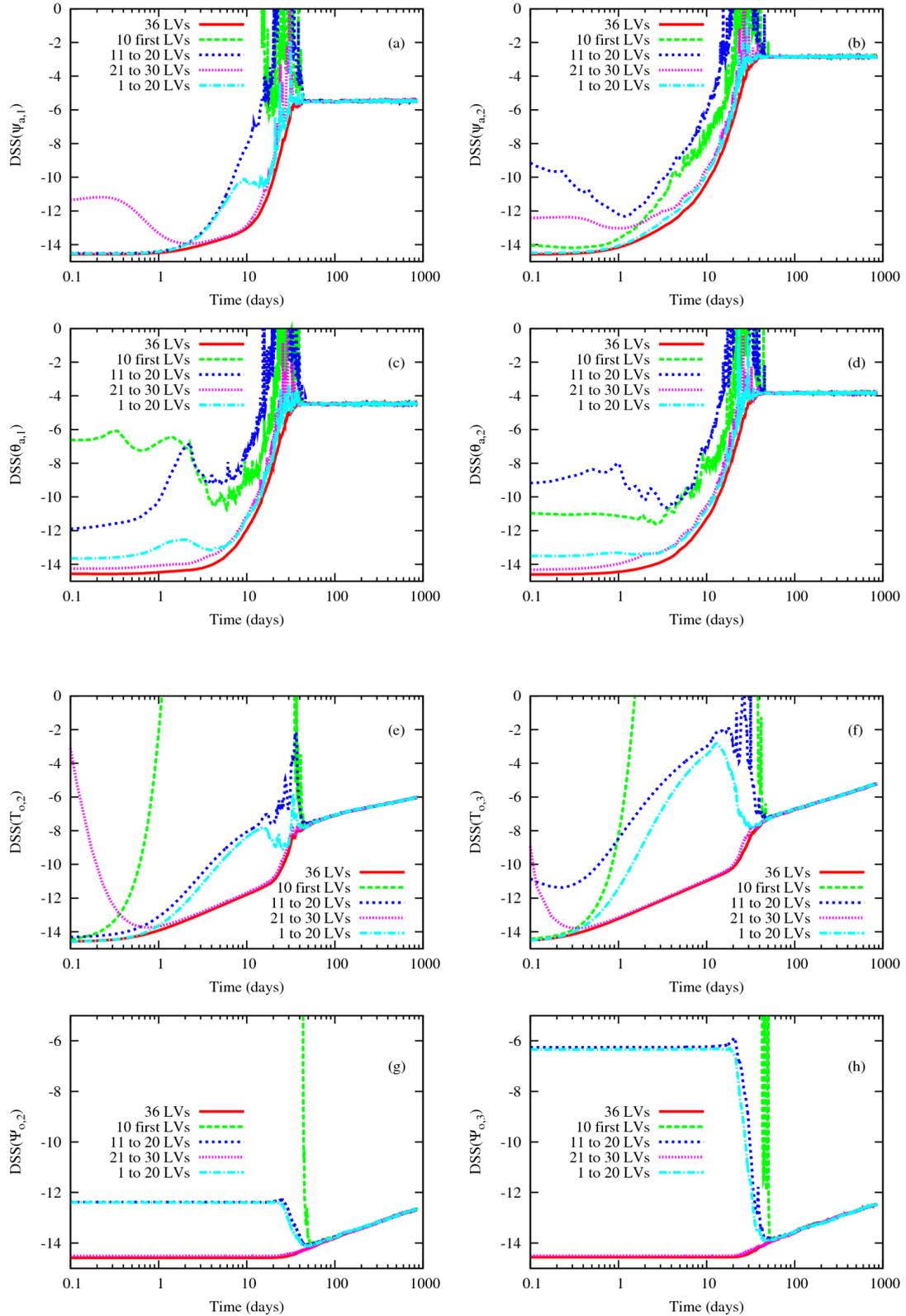

Figure 5: Dawid-Sebastiani Score (DSS) for 8 variables of the system from panel (a) to (h). In each panel the different curves represent one specific experiment. The lower the value, the better. The reference is the red continuous curve with the perfectly reliable

ensemble displayed in Fig. 3. The green dashed, the blue short-dashed, the pink dots and the light blue dash-dotted curves correspond to the experiments with the 1st to 10th BLVs, with the 11th to 20th BLVs, with the 21$^{st}$ to 30$^{th}$ BLVs and the 1$^{st}$ to 20$^{th}$ BLVs, respectively.

The DSS allows for evaluating together the quality of the first and second moment of the probabilistic forecast generated by the ensembles. Figure 5 displays DSS for 8 variables of the system, $\psi_{a,1}$, $\psi_{a,2}$, $\theta_{a,1}$, $\theta_{a,2}$, $T_{o,2}$, $T_{o,3}$, $\Psi_{o,2}$, $\Psi_{o,3}$. Four of them are dominant in the dynamics of the full system, $\psi_{a,1}$, $\theta_{a,1}$, $T_{o,2}$, $\Psi_{o,2}$. The four others are taken to illustrate the impact on the less prominent modes. Several curves are displayed for the different perturbation experiments, the reference being the red continuous curve as this corresponds to the experiment of Fig. 3 with the projections on the full set of BLVs. The score is a negative score, implying that the lower the value the better.

First an interesting observation is the convergence of all the curves toward the same value after about 50 days, indicating that for long times all experiments will provide the same forecast quality. Before that considerable differences are visible. First the perturbations along the set of BLVs 1 to 10 are not able to provide an appropriate ensemble forecast as the (green) dashed curve displays values much larger than the reference curve, except for variables $\psi_{a,2}$ and $\theta_{a,2}$. The latter two are variables at which the most unstable Lyapunov vector has a large projection (Vannitsem and Lucarini, 2016). When the second set of vectors from 11 to 20 are used, the DSS is usually closer to the (red) continuous curve, confirming the better quality of the ensemble forecasts with these type of perturbations. The combination of these two sets (1 to 20) is providing a very good result for the atmospheric variables, but does not improve much in the ocean. Interestingly, the best set of BLVs are the 21 to 30 that are providing the best results for all the variables, except at very short times up to a few days.

Overall, this analysis confirms the conclusions drawn with the comparison of the MSE and SPREAD of Fig. 4. One additional important detail is the fact that all perturbation approaches provide similar results after about 50 days. This can be understood by the fact that there is no low-frequency variability developing for the parameter chosen here.

This result is of course counterintuitive as we expect to get good results with the most unstable directions as usually claimed when initializing ensemble for operational forecasts. But it should be realized that the system under investigation here is a multi-scale system and the unstable directions have large components along the fast variables (see e.g. Vannitsem and Lucarini, 2016; Penny et al, 2019). The near-neutral modes and slightly negative ones have however larger projections along the ocean variables. This implies that when perturbing along the near-neutral or slightly negative ones one introduces larger perturbation amplitudes within the ocean. On the other hand, as well known in the context of dynamical systems theory, any perturbation (except the ones exactly aligned along the CLVs) will rapidly "rotate" in phase space and align along the most unstable direction (e.g. Vautard and Legras, 1996; Trevisan and Pancotti, 1998; Kuptsov and Parlitz, 2012). This is precisely what is seen here. When perturbing in the subspace defined by the 11 to 20 BLVs or the one defined by the 21 to 30 BLVs, any perturbation that is not aligned along a specific set of CLVs will rapidly amplify along the most unstable directions describing the unstable subspace of the system, either represented by the dominant BLVs or CLVs. Note that an experiment has also been done by perturbing the set from 31 to 36 leading overall to less good performances than with the two previous sets.

So if some specific directions should be selected to perturb the system, the ones associated with the dynamics of the slow manifold is best. Here these modes correspond to the BLVs associated with near-neutral or slightly negative Lyapunov exponents.

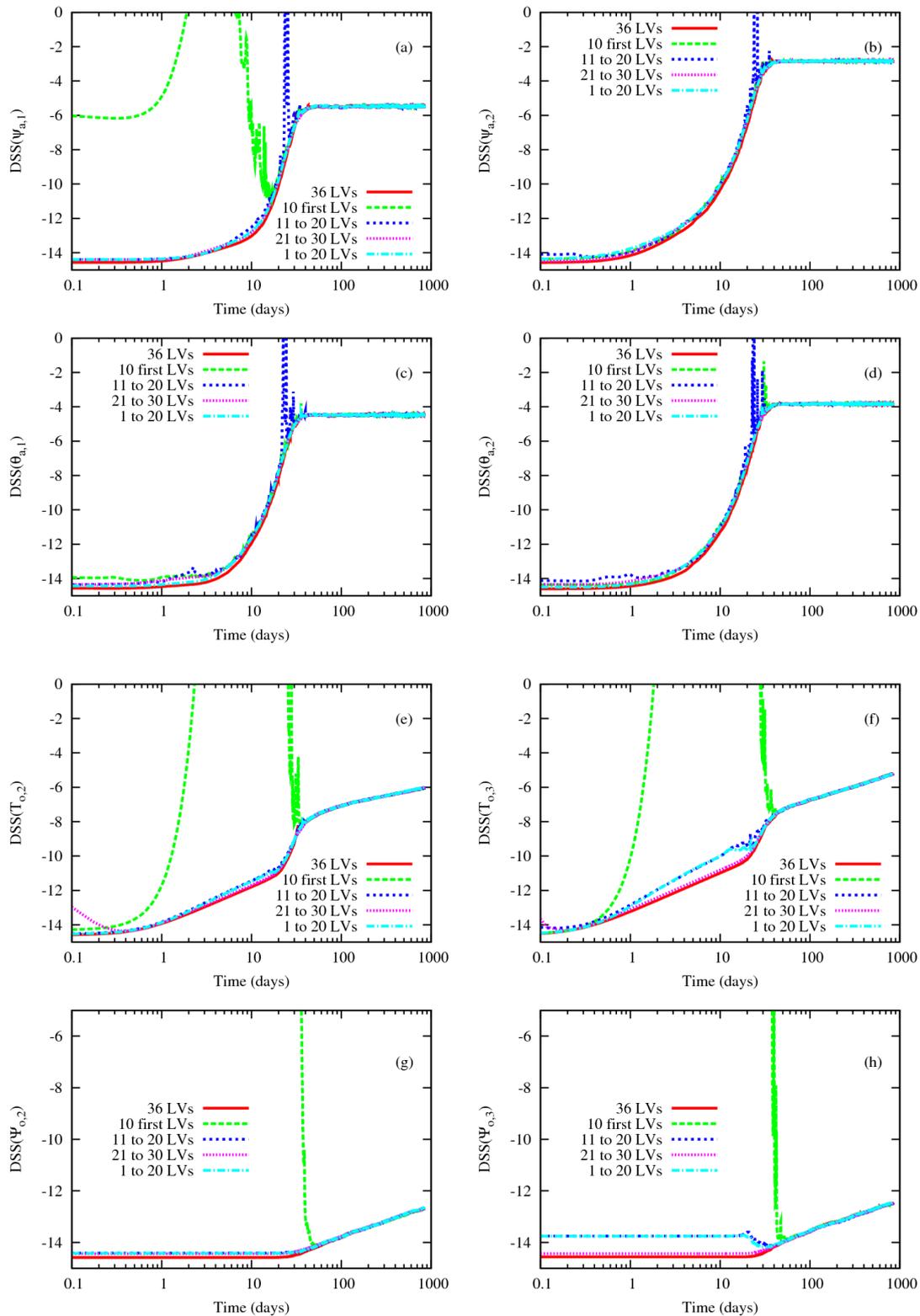

Figure 6: As in Fig. 5, but with perturbations whose amplitudes are inflated by factors 1.5, 5, 1.2 and 1.2 for the atmospheric barotropic streamfunction, the atmospheric baroclinic streamfunction, the ocean streamfunction and the ocean temperature, respectively.

We can now wonder whether by tuning the amplitudes of the perturbations along different variables of the system, and projecting along the different sets of vectors, one can get improvements. This is illustrated in Fig. 6 by increasing the amplitudes of the perturbations along the set of BLVs. Note that the increase in amplitude is different along the group of variables: (i) 1.5 times increase of amplitude for the barotropic atmospheric streamfunction; (ii) 5 times increase of amplitude for the baroclinic atmospheric streamfunction; (iii) 1.2 times increase of amplitude for both the temperature and streanfunction fields in the ocean. This corresponds to an increase of amplitude and a rotation of the perturbations in the space spanned by the sets of BLVs.

Figure 6 displays the results as in Fig. 5. Clearly the increase of amplitudes considerably improved the results with most of the different sets of BLVs. The set of dominant BLVs from 1 to 10 is still behaving very poorly for $\psi_{a,1}$ and for the ocean variables. Overall including the set of near-neutral (un)stable vectors is key in order to get accurate and reliable ensemble forecasts. This result also provides a justification to the use of Bred modes tuned to characterize the slow error growth in realistic coupled ocean-atmosphere systems in order to perform coupled ensemble forecasts (e.g. Peña and Kalnay, 2004; O'Kane et al, 2019).

Finally, the same experiments can be performed with the second set of parameters discussed in Section 2 for which a low-frequency variability is present within the atmosphere. Let us start with the comparison of the MSE and the SPREAD. The initial conditions along the trajectory are first selected based on the value of the second temperature mode $T_{o,2}$ in order to isolate contrasting situations on the attractor of the system. The threshold is fixed to $T^*_{o,2} = 0.08$ nondimensional units. If one perturbs along the set of unstable BLVs as illustrated in Fig. 7, the ensemble is even more under-dispersed when the initial conditions are taken for $T_{o,2} < 0.08$, situations for which the solution of the system is locally quite stable. Even for very long lead times the ensembles are drastically under-dispersive.

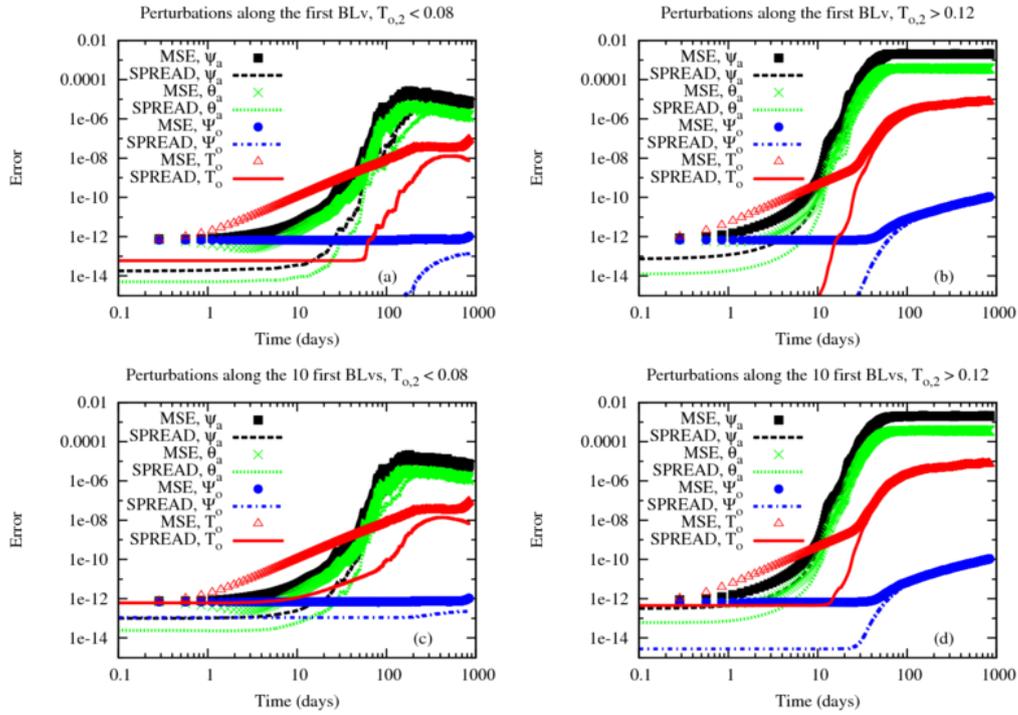

Figure 7: As in Fig. 4a and b, but now for the reference solution of Fig. 1b. The top panels are obtained with perturbations along the first BLV for different regions of the solution's attractor, namely for values of $T_{o,2} < 0.08$ (a) and $> 0.08$ (b). The bottom panels as for the top panels but with perturbations along the 10 first BLVs.

25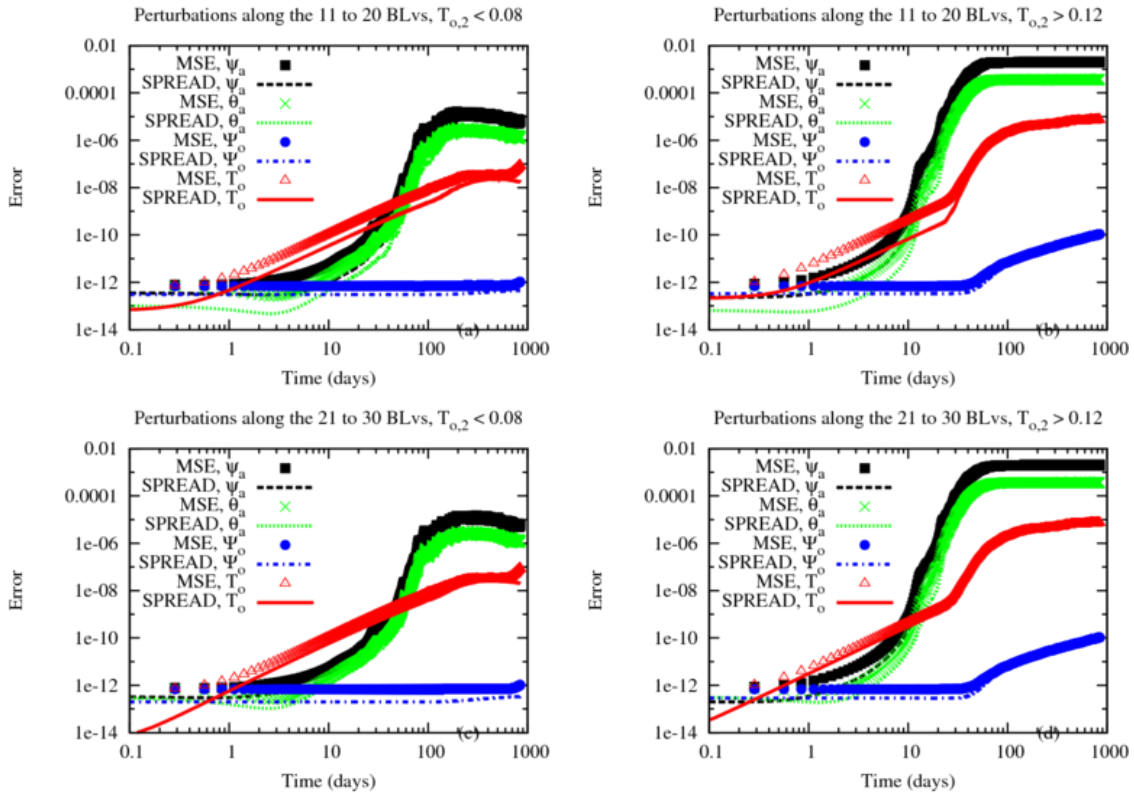

Figure 8: As in Fig. 7, but for perturbations along the near-neutral BLVs from 11 to 20, panels (a) and (b); and for perturbations along the set of BLVs from 21 to 30, panels (c) and (d).

When using the near-neutral modes (BLVs from 11 to 20) or the slightly negative ones (BLVs from 21 to 30), the ensemble forecasts are more accurate either the initial conditions are taken for values of $T_{o,2} < 0.08$, or not (Figs. 8 and 9).

Let us now look in more detail the impact of choosing different set of BLVs on the DSS. As in Fig. 5, the DSS is plotted for a selection of 8 variables (Fig. 9). Perturbations along the BLVs 1 to 10 (green dashed curve) are clearly unable to provide a DSS close to the reference (red) continuous curve of perfect ensemble reliability and accuracy at any lead times. Among the others groups of BLVs, the best result is obtained with the perturbations along the set of BLVs from 21 to 30, except at very short lead times.

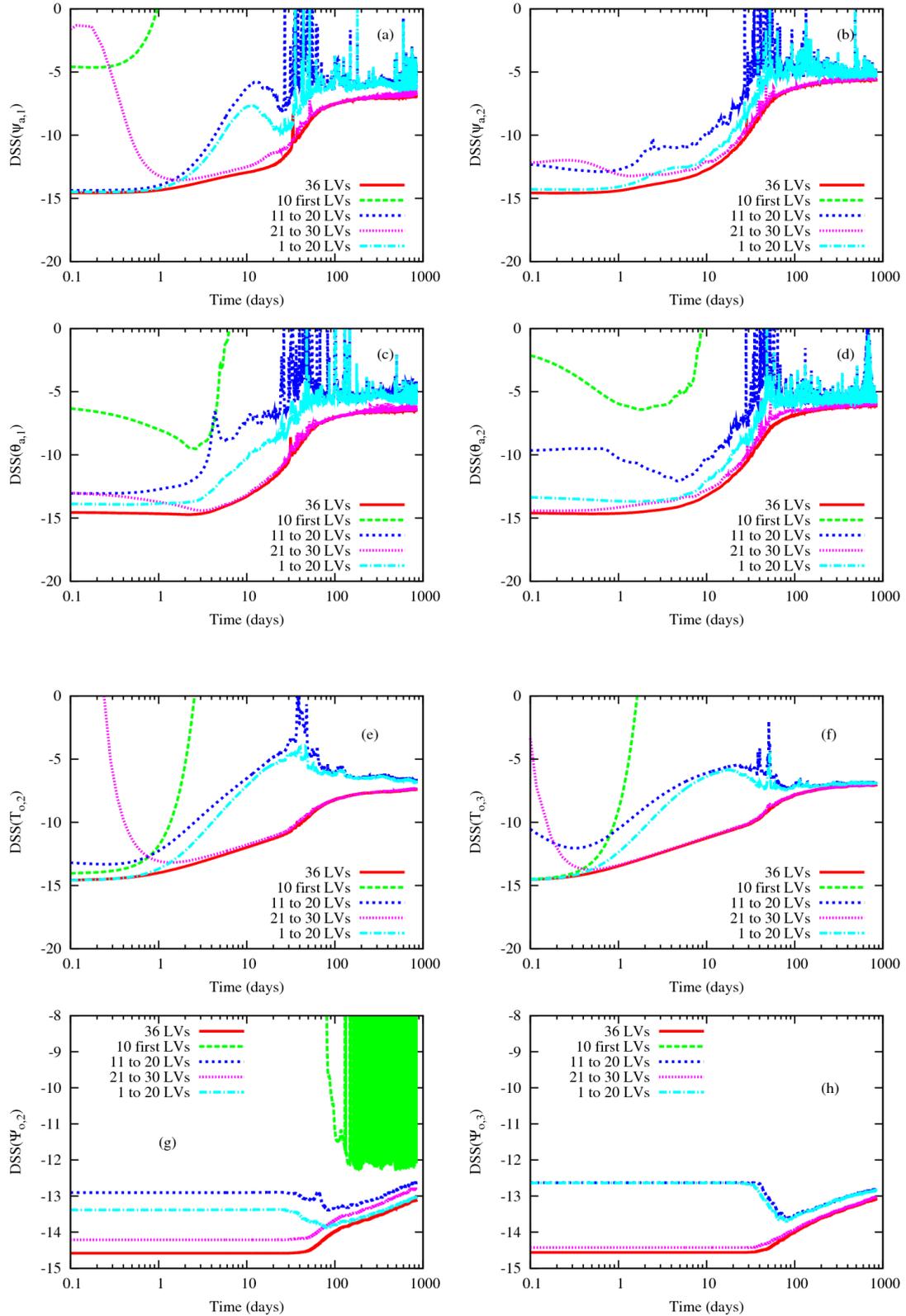

Figure 9: As in Fig. 6 but for the second set of parameters of the model. This model version displays a low-frequency variability.

It is particularly interesting to focus on the impact of this choice for long lead times. The sets of BLVs from 1 to 10, from 1 to 20, from 11 to 20 are providing results that are well above the reference, except for $\Psi_{o,2}$ (Panel (g)). The set of BLVs from 21 to 30 provide the best result with a curve almost indistinguishable from the reference.

In summary, the best set of BLVs to be used in ensemble forecasts for multiscale systems when low-frequency variability is present depends on the lead time and the observables of interest. For short times and for atmospheric variables, the most unstable ones are appropriate as also reflected in the success of ensembles for weather forecasts. For the ocean variables however these dominant modes are not the most appropriate and the near neutral ones are best. For long lead times, say from months to years, perturbing along the most unstable modes is detrimental even for atmospheric variables, and the most useful are the near neutral and slightly negative ones.

## 6. Conclusions

Ensemble forecasting in multiscale systems constitutes a new challenge for the meteorological and climate communities. One particular aspect of this problem is to define appropriate perturbations that will allow for obtaining an ensemble as reliable as possible for all components of the multiscale system. For ocean-atmosphere coupled systems, this question has been addressed by considering that slow unstable modes associated with the ocean should be perturbed in order to get information on the uncertainty of the ocean processes, and possibly for the other components of the system (e.g. Yang et al, 2009; Baehr and Piontek, 2014; O'Kane et al, 2019).

This point has been taken up here and tested in the context of reduced-order multiscale ocean-atmosphere system, known as MAOOAM, by evaluating the reliability of an ensemble forecasting system without model errors. The perturbation modes that are considered are the Backward Lyapunov Vectors known to display important similarities with Bred modes when perturbation amplitudes are small. The advantage of these vectors is that the Backward Lyapunov Vectors are independent of the rescaling time scale and amplitudes needed for defining the Bred modes, thus allowing to get more generic dynamical properties of the error behavior.

In the context of this system, it has been shown that the use of the dominant unstable modes for initialization leads to reliable forecasts for the atmospheric variables only. This reliability is further limited to short lead times when low-frequency variability is present in the coupled system. While the use of the set of near-neutral or slightly negative Lyapunov exponents provides reliability for both the atmospheric and ocean fields from medium range (weeks) to annual and decadal timescales. These modes have larger projections along the ocean variables than the others, allowing (i) for describing in a proper way the error dynamics for this component of the system and (ii) at the same time for inducing a rapid amplification of errors within the (fast) atmospheric component of the coupled system, due to the natural rotation of any perturbation toward the most unstable direction.

This very unexpected and interesting result supports the approach adopted recently of perturbing the slow unstable modes of the ocean instead of the fast scales of the atmosphere. It is however necessary to optimize in a better way the types of perturbations needed to be able to produce forecasts that are reliable for all oceanic and atmospheric variables. The present work is a contribution in that direction enlightening the role of BLVs associated with the near-neutral and slightly negative Lyapunov exponents.

29Further analyses are however necessary to extent these results to more realistic coupled ocean-atmosphere models and to select appropriate sets of modes in an optimal way. At the same time additional analyses in a hierarchy of reduced order models should be performed with a detailed comparison of the impact of using Bred modes for both the ocean and the atmosphere, and to compare these with a careful selection of BLVs (for both slow and fast time scales). Both aspects will be taken up in the near future.

Finally, one may wonder whether this type of argument could not be valid for the atmosphere itself when for instance sub-seasonal forecasting is the aim. Perturbing along BLVs associated with the planetary scale dynamics could provide more reliable and accurate sub-seasonal forecasts. This aspect will also be addressed in the future.

### *Acknowledgments.*

The authors are grateful for the very constructive comments of Steve Penny and an anonymous reviewer. The authors acknowledge the support of the Open Research Program of LASG, during which part of this work has been fulfilled. Dr. Wansuo Duan is also supported by the National Key R&D Program of China (Grant No. 2018YFC1506402) and the Natural Sciences Foundation of China (Grant No.41930971)

# REFERENCES

Baehr, J. and Piontek, R. (2014) Ensemble initialization of the oceanic component of a coupled model through bred vectors at seasonal-to-interannual timescales. Geosci. Mod Dev, **7**, 453–461.

Benedetti R (2010) Scoring rules for forecast verification. Mon Wea Rev, 138, 203-211.

Buizza R (2019) Introduction to the special issue on "25 years of ensemble forecasting". Quart J Royal Met Soc, 145, 1– 11.

Buizza, R, Leutbecher M and Isaksen, L (2008) Potential use of an ensemble of analyses in the ECMWF Ensemble Prediction System. Quart J Royal Met Soc, 134, 2051-2066.

Buizza R, Miller M and Palmer TN (1999) Stochastic representation of model uncertainties in the ECMWF ensemble prediction system. Quart J Royal Met Soc, 125, 2887-2908.

Cai M., Kalnay E and Toth Z (2003) Bred Vectors of the Zebiak–Cane Model and Their Potential Application to ENSO Predictions. J Climate, 16, 40–56.

Dawid AP and Sebastiani P (1999) Coherent dispersion criteria for optimal experimental design. The annals of Statistics, 27, 67-81.

De Cruz L, Demaeyer J and Vannitsem S (2016) The Modular Arbitrary-Order Ocean-Atmosphere Model : MAOOAM v1.0. Geosci Mod Dev, 9, 2793-2808.

Duan WS and Huo ZH (2016) An approach to generating mutually independent initial perturbations for ensemble forecasts: orthogonal conditional nonlinear optimal perturbations. J Atmos Sci, 73, 997–1014.

Epstein ES (1969) The Role of Initial Uncertainties in Predicion. J Appl Meteor, 8, 190-198.

Feng J, Ding R, Li J and Liu D (2016) Comparison of nonlinear local Lyapunov vectors with bred vectors, random perturbations and ensemble transform Kalman filter strategies in a barotropic model. Adv Atmos Sci, 33, 1036. https://doi.org/10.1007/s00376-016-6003-4

Frederiksen JS, Frederiksen CS and Osbrough SL (2010) Seasonal ensemble prediction with a coupled ocean–atmosphere model. Aust Meteor Oceanogr J, 59, 53–66, https://doi.org/10.22499/2.5901.007.

Gaspard P (1998) Chaos, Scattering and Statistical Mechanics. Cambridge University Press, UK.

Gneiting T, Balabdaoui F and Raftery A (2007) Probabilistic forecasts, calibration and sharpness. J R Statist Soc B, 69, 243-268.

Kalnay E (2003) Atmospheric modelling, data assimilation and predictability. Cambridge University Press. UK.

Kuptsov PV and Parlitz U (2012) Theory and Computation of Covariant Lyapunov Vectors. Journal of Nonlinear Science, 22, 727-762.

Legras B and Vautard R (1996) A guide to Lyapunov vectors. Seminar on Predictability, Reading, United Kingdom, ECMWF, 143–156.

Leutbecher M (2019) Ensemble size: How suboptimal is less than infinity? Quart J Royal Met Soc, 145, 107-128.

Leutbecher M and Palmer TN (2008) Ensemble Forecasting. J Comput Phys, 227, 3515-3539.

Molteni F, Buizza R, Palmer TN and Petroliagis T (1996) The ECMWF Ensemble Prediction System: Methodology and validation. Quart J Royal Met Soc, 122, 73-119. doi:10.1002/qj.49712252905

Norwood A, Kalnay E, Ide K, Yang S-C and Wolfe Ch (2013) Lyapunov, singular and bred vectors in a multi-scale system: an empirical exploration of vectors related to instabilities. J Phys A, 46, 254021.

O'Kane, TJ, Sandery PA, Monselesan DP, Sakov P, Chamberlain MA, Matear RJ, Collier MA, Squire DT, and Stevens L (2019) Coupled Data Assimilation and Ensemble Initialization with Application to Multiyear ENSO Prediction. J Climate, 32, 997–1024.

Oseledets V (2008) Oseledets Theorem, Scholarpedia, 3(1):1846.

Pazó D, Szendro IG, López JM and Rodríguez MA (2008) Structure of characteristic Lyapunov vectors in spatiotemporal chaos. Phys Rev E, 78, 016209.

Peña M and Kalnay E (2004) Separating fast and slow modes in coupled chaotic systems, Nonlin Proc Geophys, 11, 319–327, https://doi.org/10.5194/npg-11-319-2004.

Penny SG, Bach E, Bhargava K, Chang C-C, Da C, Sun L, and Yoshida, T (2019) Strongly Coupled Data Assimilation in Multiscale Media: Experiments Using a

Quasi-Geostrophic Coupled Model. Journal of Advances in Modeling Earth Systems, 11, 1803– 1829

Roulin E and Vannitsem S (2005) Skill of medium range hydrological ensemble prediction, J Hydromet, 6, 729-744.

Roulston MS and Smith LA (2002) Evaluating probabilistic forecasts using information theory. Mon Wea Rev, 130, 1653-1660.

Sandery, PA and O'Kane, TJ (2014) Coupled initialization in an ocean–atmosphere tropical cyclone prediction system. Quart J Royal Met Soc, 140, 82-95. doi:10.1002/qj.2117

Siegert S, Ferro CAT, Stephenson DB and Leutbecher M (2019) The ensemble-adjusted Ignorance Score for forecasts issued as normal distributions. Quart J Royal Met Soc, 145, 129-139.

Smith LA, Du H, Suckling EB and Niehörster F (2015) Probabilistic skill in ensemble seasonal forecasts. Quart J Royal Met Soc, 141,1085-1100.

Tebaldi C and Knutti R (2007) The use of the multi-model ensemble in probabilistic climate projections. Phil Trans Royal Soc A, 365, 2053-2075, http://doi.org/10.1098/rsta.2007.2076

Toth Z and Kalnay E (1993) Ensemble Forecasting at NMC: The Generation of Perturbations. Bull Amer Meteor Soc, 74, 2317–2330.

Trevisan A and Pancotti F (1998) Periodic Orbits, Lyapunov Vectors, and Singular Vectors in the Lorenz System. J Atmos Sci, 55, 390-398.

Vallis G (2006) Atmospheric and Oceanic Fluid Dynamics. Cambridge University Press, UK.

Vannitsem S (2015) The role of the ocean mixed layer on the development of the North Atlantic Oscillation: A dynamical system's perspective. Geophys Res Lett, 42, doi:10.1002/2015GL065974.

Vannitsem S (2017) Predictability of large-scale atmospheric motions: Lyapunov exponents and error dynamics. Chaos 27, 032101, doi: 10.1063/1.4979042

Vannitsem S, Demaeyer J, De Cruz L, Ghil M (2015) Low-frequency variability and heat transport in a low-order nonlinear coupled ocean-atmosphere model. Physica D, 309, 71-85.

Vannitsem S and Lucarini V (2016) Statistical and dynamical properties of covariant Lyapunov vectors in a coupled ocean-atmosphere model – Multiscale effects, geometric degeneracy and error dynamics, J Phys A, 49, 224001.

Vikhliaev Y, Kirtman B and Schopf P (2007) Decadal north Pacific bred vectors in a coupled GCM. J Climate, 20, 5744, doi:10.1175/2007JCLI1620.1.

Wilks DS (2011) Statistical methods in the atmospheric sciences. Academic Press.

Yang S, Kalnay E, Cai M, Rienecker M, Yuan G, and Toth Z (2006) ENSO Bred Vectors in Coupled Ocean–Atmosphere General Circulation Models. J Climate, 19, 1422–1436.

Yang S, Keppenne C, Rienecker M, and Kalnay E (2009) Application of Coupled Bred Vectors to Seasonal-to-Interannual Forecasting and Ocean Data Assimilation. J Climate, 22, 2850 –2870.

Zanna L, Brankart JM, Huber M, Leroux S, Penduff T, Williams PD (2019) Uncertainty and scale interactions in ocean ensembles: From seasonal forecasts to multidecadal climate predictions. Quart J Royal Met Soc, 145, 160–175.